\documentclass[12pt]{article}
\usepackage{amsmath}
\usepackage{graphicx}
\usepackage{amsfonts}
\usepackage{amssymb}

\newtheorem{theorem}{Theorem}
\newtheorem{definition}[theorem]{Definition}
\input{tcilatex}

\begin{document}

\title{ Towards the Construction of Wightman Functions of Integrable Quantum
Field Theories\thanks{%
To appear in the proceedings of the '6th International Workshop on Conformal
Field Theories and Integrable Models', in Chernogolovka, September 2002. }}
\author{H. Babujian \thanks{%
Permanent address: Yerevan Physics Institute, Alikhanian Brothers 2,
Yerevan, 375036 Armenia.} \thanks{
e-mail: babujian@lx2.yerphi.am, babujian@physik.fu-berlin.de}~ and M.
Karowski \thanks{
e-mail: karowski@physik.fu-berlin.de} \\
Institut f\"ur Theoretische Physik\\
Freie Universit\"at Berlin, \\
Arnimallee 14, 14195 Berlin, Germany}
\date{January 15, 2003}
\maketitle

\begin{abstract}
The purpose of the ``bootstrap program'' for integrable quantum field
theories in 1+1 dimensions is to construct a model in terms of its Wightman
functions explicitly. In this article, this program is mainly illustrated in
terms of the sine-Gordon and the sinh-Gordon model and (as an exercise) the
scaling Ising model. We review some previous results on sine-Gordon breather
form factors and quantum operator equations. The problem to sum over
intermediate states is attacked in the short distance limit of the two point
Wightman function for the sinh-Gordon and the scaling Ising model.
\end{abstract}

\section{Introduction}

The `bootstrap program for integrable quantum field theories in
1+1-dimen\-sions' \cite{STW,K,K2} does \underline{\emph{not}} start with any
classical Lagrangian. Rather this program classifies integrable quantum
field theoretic models and in addition it provides their explicit exact
solutions in term of all Wightman functions. We have contact with the
classical models only, when at the end we compare our exact results with
Feynman graph expansions which are based on the Lagrangians.

One of the authors (M.K.) et al.~\cite{KTTW} formulated the on-shell program
i.e. the exact determination of the scattering matrix using the Yang-Baxter
equations. Off-shell quantities, namely form factors were first investigated
by Vergeles and Gryanik \cite{VG} in the sinh-Gordon model and by Weisz \cite%
{W} in the sine-Gordon model. The concept of generalized form factors was
introduced by one of the authors (M.K.) et al.~\cite{KW}. In this article
consistency equations were formulated which are expected to be satisfied by
these objects. Thereafter this approach was developed further and studied in
the context of several explicit models by Smirnov \cite{Sm} who proposed the
form factor equations $(i)-(v)$ (see below) as extensions of similar
formulae in the original article \cite{KW}. The formulae were proven by the
authors et al.~\cite{BFKZ}.

Finally the Wightman functions are obtained by taking integrals and sums
over intermediate states. The explicit evaluation of all these integrals and
sums remains an open challenge for almost all models, except the Ising model 
\cite{B-J,BKW,YZ}. In this article we attack this problem for the examples
of the sinh-Gordon model and, as an exercise, of the scaling Ising model. We
investigate the short distance behavior of the two-point Wightman function
of the exponential of the field.

\section{The ``bootstrap program''}

As the final result the `bootstrap program' for integrable quantum field
theories in 1+1-dimensions provides a model in term of all Wightman
functions. The result is obtained in three steps:

\begin{enumerate}
\item The S-matrix is calculated by means of general properties such as
unitarity and crossing, the Yang-Baxter equations (which are a consequence
of integrability) and the additional assumption of `maximal analyticity'.
This means that the two-particle S-matrix is an analytic function in the
physical plane (of the Mandelstam variable $(p_{1}+p_{2})^{2}$) and
possesses only those poles there which are of physical origin. The only
input which depends on the model is the assumption of a particle spectrum.
Usually it belongs to representations of a symmetry. Typically there is a
correspondence of fundamental representations with multiplets of particles.
A \emph{classification} of all S-matrices obeying the given properties is
obtained.

\item Generalized form factors which are matrix elements of local operators 
\begin{equation*}
^{out}\left\langle \,p_{m}^{\prime},\ldots,p_{1}^{\prime}\left| \mathcal{O}%
(x)\right| p_{1},\ldots,p_{n}\,\right\rangle ^{in}\, 
\end{equation*}
are calculated by means of the S-matrix. More precisely, the equations $%
(i)-(v)$ as listed in section \ref{s3} are solved. These equations follow
from LSZ-assumptions and again the additional assumption of `maximal
analyticity' \cite{BFKZ}.

\item The Wightman functions are obtained by inserting a complete set of
intermediate states. In particular the two point function for a hermitian
operator $\mathcal{O}(x)$ reads 
\begin{multline*}
\langle\,0\left| \mathcal{O}(x)\,\mathcal{O}(0)\right| 0\,\rangle=\sum
_{n=0}^{\infty}\frac{1}{n!}\idotsint\frac{dp_{1}\ldots dp_{n}}{\,(2\pi
)^{n}2\omega_{1}\dots2\omega_{n}} \\
\times\left| \left\langle \,0\left| \mathcal{O}(0)\right|
p_{1},\ldots,p_{n}\,\right\rangle ^{in}\right| ^{2}e^{-ix\sum p_{i}}.
\end{multline*}
Up to now a proof that these sums converge exists only for the scaling Ising
model \cite{B-J} and the `Yang-Lee' model \cite{Sm3}.
\end{enumerate}

\subsection*{Integrability}

Integrability in (quantum) field theories means that there exist $\infty$%
-many local conservation laws 
\begin{equation*}
\partial_{\mu}J_{L}^{\mu}(t,x)=0\quad(L=\pm1,\pm3,\dots)\,. 
\end{equation*}
A consequence of such conservation laws in 1+1 dimensions is that there is
no particle production and the n-particle S-matrix is a product of
2-particle S-matrices 
\begin{equation*}
S^{(n)}(p_{1},\dots,p_{n})=\prod_{i<j}S_{ij}(p_{i},p_{j})\,. 
\end{equation*}
If backward scattering occurs the 2-particle S-matrices will not commute and
one has to specify the order. In particular for the 3-particle S-matrix
there are two possibilities 
\begin{gather*}
\mathbf{S^{(3)}=S_{12}S_{13}S_{23}=S_{23}S_{13}S_{12}} \\[2mm]
\begin{array}{c}
\unitlength5mm\begin{picture}(17,4)(-6,0) \thicklines\put(-5.5,.8){\line
(1,1){3}} \put(-4,.8){\line(0,1){3}}
\put(-2.5,.8){\line(-1,1){3}}\put(-4,2.3){\makebox(0,0){\Large$\bullet$}}
\put(-5.6,0){$1$} \put (-4.2,0){$2$} \put(-2.8,0){$3$} \put(-1.5,2){$=$}
\put(0,1){\line(1,1){3}}\put(0,3){\line(1,-1){3}} \put(2,0){\line(0,1){4}}
\put(4.3,2){$=$}\put(6,0){\line(1,1){3}} \put(6,4){\line(1,-1){3}}
\put(7,0){\line(0,1){4}}\put(.2,.5){$1$} \put(1.3,0){$2$} \put(3,.2){$3$}
\put(5.5,.2){$1$}\put(7.3,0){$2$} \put(8.4,.4){$3$} \end{picture}%
\end{array}%
\end{gather*}
which yield the \textbf{``Yang-Baxter Equation''.}\\[3pt]
\textbf{Examples }of integrable models in 1+1-dimensions are the \textbf{%
sine-Gordon} model defined by the classical field equation 
\begin{equation*}
\ddot{\varphi}(t,x)-\varphi^{\prime\prime}(t,x)+\frac{\alpha}{\beta}\sin
\beta\varphi(t,x)=0 
\end{equation*}
and the massive Thirring model defined by the classical Lagrangian 
\begin{equation*}
\mathcal{L}=\bar{\psi}(i\gamma\partial-m)\psi-\tfrac{1}{2}\,g\,\,\bar{\psi }%
\gamma^{\mu}\psi\,\bar{\psi}\gamma_{\mu}\psi\,. 
\end{equation*}
Coleman \cite{Co} proved that both models are equivalent on the quantum
level.

Further integrable models are: $Z_{N}$-Ising models, nonlinear $\sigma $%
-models, Gross-Neveu models, Toda models etc. In the following most of the
formulae and explicit solutions are given for the sine-Gordon alias massive
Thirring model although often corresponding results exist also for other
models.

\subsection*{The S-matrix}

For the Sine-Gordon alias massive Thiring model the particle spectrum
consists of: soliton, anti-soliton and breathers (as soliton anti-soliton
bound states). Since backward scattering can only appear for particles with
the same mass, the two-particle S-matrix is of the form 
\begin{equation*}
S(\theta_{12})=\left( 
\begin{array}{ccccccc}
u &  &  &  &  &  &  \\ 
& t & r &  &  &  &  \\ 
& r & t &  &  &  &  \\ 
&  &  & u &  &  &  \\ 
&  &  &  & S_{sb} &  &  \\ 
&  &  &  &  & S_{bb} &  \\ 
&  &  &  &  &  & \ddots%
\end{array}
\right) 
\end{equation*}
where the rapidity difference $\theta_{12}=|\theta_{1}-\theta_{2}|$ is
defined by $p_{i}=m_{i}(\cosh\theta_{i},\sinh\theta_{i})$.

We start with the soliton (anti-soliton) S-matrix: 
\begin{equation*}
S_{\alpha\;\beta}^{\beta^{\prime}\alpha^{\prime}}=%
\begin{array}{c}
\unitlength2mm\begin{picture}(5,6.3)
\thicklines\put(0,1){\line(1,1){4}}\put(2,3){\makebox(0,0){$\bullet$}}
\put(4,1){\line(-1,1){4}} \put (0,-.5){$\alpha$} \put(3.5,-.5){$\beta$}
\put(3.5,5.8){$\alpha'$}\put(0,5.8){$\beta'$} \end{picture}%
\end{array}
:\quad S_{ss}^{ss}=u,\quad S_{s\bar{s}}^{\bar{s}s}=t,\quad S_{s\bar{s}}^{s%
\bar{s}}=r 
\end{equation*}
$s=$ soliton, $\bar{s}=$ anti-soliton. As input conditions we have:

\begin{enumerate}
\item {Unitarity:} $S(-\theta)S(\theta)=1$ 
\begin{align*}
u(-\theta)u(\theta) & =1 \\
t(-\theta)t(\theta)+r(-\theta)r(\theta) & =1 \\
t(-\theta)r(\theta)+r(-\theta)t(\theta) & =0
\end{align*}

\item {Crossing:} 
\begin{equation*}
u(i\pi-\theta)=t(\theta)~,~~r(i\pi-\theta)=r(\theta) 
\end{equation*}

\item {Yang-Baxter:} 
\begin{equation*}
r(\theta_{12})u(\theta_{13})r(\theta_{23})+t(\theta_{12})r(\theta
_{13})t(\theta_{23})=u(\theta_{12})r(\theta_{13})u(\theta_{23}) 
\end{equation*}

\item \emph{Maximal analyticity:}

$S(\theta)$ is meromorphic in the `physical strip' $0\leq\func{Im}%
\theta\leq\pi$ and all poles there have a physical interpretation, in
particular all bound states correspond to simple poles. An S-matrix
satisfying this condition is also called `minimal'. For couplings $g<0$ (in
the language of the massive Thirring model) there are no soliton
anti-soliton bound states. Therefore in this region of the coupling constant 
$S(\theta)$ is holomorphic in the `physical strip' $0\leq\func{Im}%
\theta\leq\pi$.
\end{enumerate}

The S-matrix bootstrap using the {Yang-Baxter} relations was proposed by
Karowski, Thun, Truong and Weisz \cite{KTTW}. It was shown in this article
that the 'minimal' {general solution} to these equations is 
\begin{equation*}
u(\theta,\nu)=\exp\int_{0}^{\infty}\frac{dt}{t}\,\frac{\sinh\frac{t}{2}%
(1-\nu)}{\sinh\frac{\nu t}{2}\,\cosh\frac{t}{2}}\,\sinh t\frac{\theta}{i\pi }%
\,. 
\end{equation*}
This S-matrix was first obtained by Zamolodchikov \cite{Za} from the
extrapolation of semi-classical expressions. It has been checked in
perturbation theory. The free parameter $\nu$ is related to the coupling
constants by 
\begin{equation*}
\frac{1}{\nu}=\frac{8\pi}{\beta^{2}}-1=1+\frac{2g}{\pi}\,. 
\end{equation*}
The second equation is due to Coleman and the first one is obtained by
analyzing the pole structure of the amplitude $u(\theta,\nu)$. The
assumption of 'maximal analyticity' and comparison with the known
semi-classical bound state spectrum provides the identification of the
parameter $\nu$.

The two-breather S-matrix 
\begin{equation}
S_{bb}(\theta)=\frac{\sinh\theta+i\sin\pi\nu}{\sinh\theta-i\sin\pi\nu} 
\label{2.1}
\end{equation}
is obtained by the bound state fusion method \cite{KT}.

{\ }

\section{Form factors\label{s3}}

\begin{definition}
For a local operator $\mathcal{O}(x)$ the generalized form factors \cite{KW}
are defined as 
\begin{equation*}
\,\mathcal{O}_{\alpha_{1}\dots\alpha_{n}}\left( \theta_{1},\dots,\theta
_{n}\right) =\langle\,0\,|\,\mathcal{O}(0)\,|\,p_{1},\dots,p_{n}\,\rangle_{%
\alpha_{1}\dots\alpha_{n}}^{in}
\end{equation*}
for $\theta_{1}>\dots>\theta_{n}$. For other orders of the rapidities they
are defined by analytic continuation. The index $\alpha_{i}$\thinspace
denotes the type of the particle with momentum $p_{i}$. We also use the
short notations $\mathcal{O}_{\underline{\alpha}}(\underline{\theta})$ or $%
\mathcal{O}_{1\dots n}(\underline{\theta})$.
\end{definition}

For the sine Gordon model $\alpha$ denotes the soliton, anti-soliton or
breathers. Similar as for the S-matrix, `maximal analyticity' for
generalized form factors means again that they are meromorphic and all poles
in the `physical strips' $0\leq\func{Im}\theta_{i}\leq\pi$ have a physical
interpretation. Together with the usual LSZ-assumptions \cite{LSZ} of local
quantum field theory the following form factor equations can be derived

\begin{enumerate}
\item[$(i)$] Watson's equations \cite{Wa}: 
\begin{equation}
\mathcal{O}_{\dots ij\dots}(\dots,\theta_{i},\theta_{j},\dots)=\mathcal{O}%
_{\dots
ji\dots}(\dots,\theta_{j},\theta_{i},\dots)\,S_{ij}(\theta_{i}-\theta_{j})%
\,.   \label{pf1}
\end{equation}

\item[$(ii)$] Crossing relation (for the connected part of the matrix
element): 
\begin{align}
_{\bar{\alpha}_{1}}\langle\,p_{1}\,|\,\mathcal{O}(0)\,|\,p_{2},\dots
,p_{n}\,\rangle_{\alpha_{2}\dots\alpha_{n}}^{in,conn.} & =\mathcal{O}%
_{\alpha_{1}\alpha_{2}\dots\alpha_{n}}(\theta_{1}+i\pi,\theta_{2},\dots
,\theta_{n})  \label{pf2} \\
& =\mathcal{O}_{\alpha_{2}\dots\alpha_{n}\alpha_{1}}(\theta_{2},\dots
,\theta_{n},\theta_{1}-i\pi)\,.  \notag
\end{align}

\item[$(iii)$] Recursion relation: 
\begin{equation}
\limfunc{Res}_{\theta_{12}=i\pi}\mathcal{O}_{1\dots n}(\theta_{1},\dots)=2i\,%
\mathbf{C}_{12}\,\,\mathcal{O}_{3\dots n}(\theta_{3},\dots)\left( \mathbf{1}%
-S_{2n}\dots S_{23}\right) \,,   \label{pf3}
\end{equation}
where $\mathbf{C}_{12}$ is the charge conjugation matrix.

\item[$(iv)$] Bound state form factors equations: 
\begin{equation}
\limfunc{Res}_{\theta_{12}=ia}\mathcal{O}_{123\dots n}(\underline {\theta})=%
\mathcal{O}_{(12)3\dots n}(\theta_{(12)},\underline{\theta}^{\prime })\,%
\sqrt{2}\,\Gamma_{12}^{(12)}\,,   \label{pf4}
\end{equation}
where $a$ is the fusion angle and $\Gamma_{12}^{(12)}$ is the fusion
intertwiner \cite{K1,BK1}.

\item[$(v)$] Lorentz invariance: 
\begin{equation}
\mathcal{O}_{1\dots n}(\theta_{1}+u,\dots,\theta_{n}+u)=e^{su}\,\mathcal{O}%
_{1\dots n}(\theta_{1},\dots,\theta_{n})\,,   \label{pf5}
\end{equation}
where $s$ is the ``spin'' of $\mathcal{O}$.
\end{enumerate}

These equations have been proposed by Smirnov \cite{Sm} as generalizations
of equations derived in the original articles \cite{KW,BKW,K2}. They have
been proven \cite{BFKZ} by means of the LSZ-assumptions and `maximal
analyticity'. They hold in this form for bosons; for fermions or more
generally for anyons there are some additional phase factors.

\subsection*{Two-particle form factors}

For the two-particle form factors the form factor equations are easily
understood. The usual assumptions of local quantum field theory yield 
\begin{equation*}
\langle\,0\,|\,\mathcal{O}(0)\,|\,p_{1},p_{2}\rangle^{in/out}=F\left(
(p_{1}+p_{2})^{2}\pm i\varepsilon\right) =\,F\left( \pm\theta_{12}\right) 
\end{equation*}
where the rapidity difference is defined by $p_{1}p_{2}=m^{2}\cosh\theta_{12}
$. For integrable theories one has particle number conservation which
implies (for any eigenstate of the two-particle S-matrix) 
\begin{equation*}
\langle\,0\,|\,\mathcal{O}(0)\,|\,p_{1},p_{2}\rangle^{in}=\langle \,0\,|\,%
\mathcal{O}(0)\,|\,p_{2},p_{1}\rangle^{out}\,S\left( \theta _{12}\right) \,. 
\end{equation*}
Crossing (\ref{pf2}) means 
\begin{equation*}
\langle\,p_{1}\,|\,\mathcal{O}(0)\,|\,p_{2}\rangle=F\left( i\pi-\theta
_{12}\right) 
\end{equation*}
where for one-particle states in- and out-states coincide. Therefore
Watson's equations follow 
\begin{equation*}
\begin{array}{l}
F\left( \theta\right) =F\left( -\theta\right) S\left( \theta\right) \vspace{%
5pt} \\ 
F\left( i\pi-\theta\right) =F\left( i\pi+\theta\right) \,.%
\end{array}
\end{equation*}
For general theories Watson's \cite{Wa} equations only hold below the
particle production thresholds. However, for integrable theories there is no
particle production and therefore they hold for all complex values of $\theta
$. It has been shown \cite{KW} that these equations together with ``maximal
analyticity'' have a unique solution.

As an example we write the sine-Gordon two-breather form factor \cite{KW} 
\begin{equation}
F(\theta)=\exp\int_{0}^{\infty}\frac{dt}{t\sinh t}\,\left( \frac{\cosh (%
\frac{1}{2}+\nu)t}{\cosh\frac{1}{2}t}-1\right) \cosh t\left( 1-\frac {\theta%
}{i\pi}\right)   \label{3.6}
\end{equation}

\subsection*{A formula for sine-Gordon breather form factors}

We are looking for solutions of the form factor equations $(i)-(v)$. The
generalized form factors for arbitrary numbers of breathers are of the form 
\cite{KW} 
\begin{equation}
\langle\,0\,|\,\mathcal{O}(0)\,|p_{1},\ldots,p_{n}\,\rangle^{in}=K_{n}^{%
\mathcal{O}}(\underline{\theta})\prod_{1\leq i<j\leq n}F(\theta _{ij}) 
\label{2}
\end{equation}
where $F(\theta)$ is the two-breather form factor above and the `K-function'
satisfies Watsons equations with $S=1$. We make the \emph{Ansatz}\footnote{%
Using an integral representation \cite{BFKZ,BK} for general soliton form
factors we derived this formula in \cite{BK} for several local operators
(see also \cite{Sm2}). Here we consider it as an Ansatz also for operators
such as the general exponential of the breather field which is nonlocal with
respect to the soliton field.}%
\begin{equation}
K_{n}^{\mathcal{O}}(\underline{\theta})=\sum_{l_{1}=0}^{1}\dots%
\sum_{l_{n}=0}^{1}(-1)^{\sum l_{i}}\prod_{i<j}\!\left( 1+(l_{i}-l_{j})\frac{%
i\sin\pi\nu }{\sinh\theta_{ij}}\right) \!p_{n}^{\mathcal{O}}(\underline{%
\theta },\underline{l})\,.   \label{3}
\end{equation}
The dependence of form factors on the operator $\mathcal{O}(x)$ enters only
through the p-functions $p_{n}^{\mathcal{O}}(\underline{\theta},{\underline {%
z})}$. This Ansatz transforms the form factor equations $(i)-(v)$ to simpler
equations for the p-function $p_{n}^{\mathcal{O}}(\underline{\theta },%
\underline{l})$ \cite{BK2}. The p-function $p_{n}^{\mathcal{O}}(\underline{%
\theta},\underline{l})$ is holomorphic with respect to all variables $%
\theta_{1},\dots,\theta_{n}$. It is symmetric with respect to the exchange
of the variables $\theta_{i}$ and $l_{i}$ at the same time and it is
periodic with period $2\pi i$. 
\begin{align}
p_{n}^{\mathcal{O}}(\dots,\theta_{i},\theta_{j},\dots,l_{i},l_{j},\dots) &
=p_{n}^{\mathcal{O}}(\dots,\theta_{j},\theta_{i},\dots,l_{j},l_{i},\dots)
\label{pp1} \\
p_{n}^{\mathcal{O}}(\underline{\theta},\underline{l}) & =p_{n}^{\mathcal{O}%
}(\theta_{1}-2\pi i,\theta_{2},\dots,\theta_{n},\underline{l})   \label{pp2}
\end{align}
With the short hand notation $\underline{\theta}^{\prime}=\theta_{2},\dots,%
\theta_{n},\,\underline{\theta}^{\prime\prime}=\theta_{3},\dots ,\theta_{n}$
and $\underline{l}^{\prime\prime}=l_{3},\dots,l_{n}$ the recursion relation 
\begin{equation}
p_{n}^{\mathcal{O}}(\theta_{2}+i\pi,\underline{\theta}^{\prime},\underline {l%
})=g(l_{1},l_{2})p_{n-2}^{\mathcal{O}}(\underline{\theta}^{\prime\prime },%
\underline{l}^{\prime\prime})+h(l_{1},l_{2})   \label{pp3}
\end{equation}
holds where $g(0,1)=g(1,0)=2/(F(i\pi)\sin\pi\nu)$ and $h(l_{1},l_{2})$ is
independent of $\underline{l}^{\prime\prime}$. Lorentz covariance reads as 
\begin{equation}
p_{n}^{\mathcal{O}}(\theta_{1}+\mu,\dots,\theta_{n}+\mu,\underline{l}%
)=e^{s\mu}\,p_{n}^{\mathcal{O}}(\theta_{1},\dots,\theta_{n},\underline {l}). 
\label{pp5}
\end{equation}
These conditions of the p-function are sufficient to guarantee the
properties of the form factors.

\begin{theorem}
\label{t1}If the p-function $p_{n}^{\mathcal{O}}(\underline{\theta},%
\underline{l})$ satisfies the conditions (\ref{pp1}--\ref{pp5}) the form
factor function $\mathcal{O}_{n}(\underline{\theta})$ satisfies the
properties (\ref{pf1}--\ref{pf5}).
\end{theorem}

This theorem has been proven in \cite{BK2}.

\paragraph{\textbf{Examples of operators and their p-functions: }}

For several cases the correspondence between local operators and their
p-functions have been proposed in \cite{BK2}. Here we provide three examples:

\begin{enumerate}
\item The {normal ordered exponential of the field (see also \cite{BL})} 
\begin{equation}
\mathcal{O}(x)=\,:\!e^{i\gamma\varphi(x)}\!:\quad\leftrightarrow p(%
\underline{\theta},\underline{l})=\left( \frac{2}{F(i\pi)\sin\pi\nu }\right)
^{\frac{n}{2}}\prod\limits_{i=1}^{n}e^{i\pi\nu\frac{\gamma}{\beta }%
(-1)^{l_{i}}}   \label{pq}
\end{equation}

\item Expanding the last relation with respect to $\gamma$ one obtains the
p-func\-tions for {normal ordered powers }$:\!\varphi(x)^{N}\!:$ in
particular for $N=1$%
\begin{equation*}
\,:\!\varphi(x)\!:\quad\leftrightarrow p(\underline{\theta},\underline {l})=%
\frac{\pi\nu}{\beta}\left( \frac{2}{F(i\pi)\sin\pi\nu}\right) ^{\frac{n}{2}%
}\sum_{i=1}^{n}(-1)^{l_{i}}
\end{equation*}
which yields (for $n=1$) the `wave function renormalization constant' 
\begin{equation*}
Z^{\varphi}=\langle\,0\,|\,\varphi(0)\,|p\,\rangle^{2}=\frac{8\pi^{2}\nu^{2}%
}{F(i\pi)\beta^{2}\sin\pi\nu}
\end{equation*}
(see also \cite{KW}).

\item The {higher conserved currents} (which are typical for integrable
quantum field theories) 
\begin{equation*}
J_{L}^{\pm}(x)\leftrightarrow\pm N_{n}^{(J_{L})}\sum_{i=1}^{n}e^{\pm\theta
_{i}}\sum_{i=1}^{n}e^{L\left( \theta_{i}-\frac{i\pi}{2}(1-(-1)^{l_{i}}\nu)%
\right) }\,. 
\end{equation*}
\end{enumerate}

\subsection*{Asymptotic behavior of the form factors for $\,:\!e^{i\protect%
\gamma \protect\varphi}\!:$}

Let $\mathcal{O}=\,:\!\varphi^{N}\!:$ be the normal ordered power of a
bosonic field. Write the rapidities as $\underline{\theta}%
=\lambda\theta_{1}^{\prime
},\dots,\lambda\theta_{m}^{\prime},\theta_{1}^{\prime\prime},\dots
,\theta_{n-m}^{\prime\prime}$ and consider the limit $\lambda\rightarrow
\infty$. Then the asymptotic behavior of the n-boson form factor is 
\begin{align*}
\left[ \varphi^{N}\right] _{n}(\underline{\theta}) & =\langle
\,0\,|\,:\!\varphi^{N}\!:(0)\,|p_{1},\ldots,p_{n}\,\rangle^{in} \\
& =\sum_{K=0}^{N}\binom{N}{K}\left[ \varphi^{K}\right] _{m}(\underline {%
\theta}^{\prime})\,\left[ \varphi^{N-K}\right] _{n-m}(\underline{\theta }%
^{\prime\prime})+O(e^{-\lambda})\,.
\end{align*}
This can be proven in any order of perturbation theory as follows. The
matrix element on the left hand side may be written in terms of Feynman
graphs as 
\begin{equation*}
\begin{array}{c}
{\unitlength3.6mm%
\begin{picture}(6,6) \put(3,2){\oval(6,2)} \put (3,5){\makebox(0,0){$\bullet$}} \put(3,5.8){\makebox(0,0){${\cal
O}=:\varphi^N:$}} \put(1,3){\line(1,1){2}} \put(2,3){\line(1,2){1}}\put(5,3){\line(-1,1){2}} \put(3,.5){$\dots$} \put(2.7,3.5){$N$}\put(1,0){\line(0,1){1}} \put(2,0){\line(0,1){1}} \put(5,0){\line(0,1){1}}\put(-2.1,0){$\theta_1+\lambda$} \put(5.4,0){$\theta_r$} \end{picture}%
}%
\end{array}
~=\sum_{K=0}^{N}\binom{N}{K}~%
\begin{array}{c}
\unitlength3.6mm%
\begin{picture}(11,6) \put(2,2){\oval(4,2)} \put (8,2){\oval(4,2)} \put(5,5){\makebox(0,0){$\bullet$}} \put(5,5.8){\makebox
(0,0){${\cal O}=:\varphi^N:$}} \put(1,3){\line(2,1){4}} \put(3,3){\line (1,1){2}} \put(7,3){\line(-1,1){2}} \put(9,3){\line(-2,1){4}} \put (1.5,.5){$\dots$} \put(7.5,.5){$\dots$} \put(1.7,4){$K$} \put(8,4){$N-K$}\put(1,0){\line(0,1){1}} \put(3,0){\line(0,1){1}} \put(7,0){\line(0,1){1}}\put(9,0){\line(0,1){1}} \put(-.1,0){$\theta_1$} \put(3.4,0){$\theta_s$}\put(9.4,0){$\theta_r$} \end{picture}%
\end{array}
+\dots 
\end{equation*}
where all other graphs not drawn have lines which connect both parts
directly. Weinberg's power counting theorem for bosonic Feynman graphs
implies that these contributions decrease for $\lambda\rightarrow\infty$ as $%
O(\lambda ^{k}e^{-\lambda})\,$. This behavior is also assumed to hold for
the exact form factors (the fact is that the `logarithmic terms' $\lambda^{k}
$ do not show up for the exact expressions since the K-functions are
meromorphic in the $e^{\theta_{i}}$). Therefore for the exponentials of the
boson field $:e^{i\gamma\varphi}:$ we have the asymptotic behavior 
\begin{equation*}
\left[ e^{i\gamma\varphi}\right] _{n}(\underline{\theta})=\left[
e^{i\gamma\varphi}\right] _{m}(\underline{\theta}^{\prime})\,\left[
e^{i\gamma\varphi}\right] _{n-m}(\underline{\theta}^{\prime\prime
})+O(e^{-\lambda})\,. 
\end{equation*}
It is easy to see \cite{BK1,BK2} that our proposal (\ref{pq}) together with (%
\ref{2}) and (\ref{3}) satisfies this asymptotic behavior\footnote{%
This type of arguments has been also used before \cite{KW,FMS,KM,MS}.}. The
asymptotic behavior of other form factors is more complicated \cite{BK} in
particular if fermions are involved.

\subsection*{Quantum field operator equations}

\subsubsection*{The sine-Gordon equation}

We start with the local operator $:\!\sin\gamma\varphi\!:(x)=\frac{1}{2i}%
:\!\left( e^{i\gamma\varphi}-e^{-i\gamma\varphi}\right) \!:(x)$. For the
exceptional value $\gamma=\beta$ we find \cite{BK,BK2} that also $\square
^{-1}\!:\!\sin\beta\varphi\!:(x)$ is local. Moreover the quantum sine-Gordon
field equation 
\begin{equation}
\square\varphi(x)+\frac{\alpha}{\beta}:\!\sin\beta\varphi\!:(x)=0   \label{e}
\end{equation}
holds for all matrix elements, if the ``bare'' mass $\sqrt{\alpha}$ is
related to the renormalized mass by\footnote{%
Before such a formula was found \cite{Fa,Zl1} by different methods.} 
\begin{equation}
\alpha=m^{2}\frac{\pi\nu}{\sin\pi\nu}   \label{mass}
\end{equation}
where $m$ is the physical mass of the fundamental boson. The result may be
checked in perturbation theory by Feynman graph expansions. In particular in
lowest order the relation between the bare and the renormalized mass (\ref%
{mass}) had already been calculated in the original article \cite{KW}. The
result is 
\begin{equation*}
m^{2}=\alpha\left( 1-\frac{1}{6}\left( \frac{\beta^{2}}{8}\right)
^{2}+O(\beta^{6})\right) 
\end{equation*}
which agrees with the exact formula above.

Here is a sketch of the proof of the field equation \cite{BK2} which uses
induction and Liouville's theorem. Consider the K-functions of the left hand
side of eq.~(\ref{e}) 
\begin{equation*}
f_{n}(\underline{\theta})=-\sum e^{\theta_{i}}\sum
e^{-\theta_{i}}K_{n}^{(1)}(\underline{\theta})+\frac{\pi\nu}{\beta\sin\pi\nu}%
\frac{1}{2i}\left( K_{n}^{(q)}(\underline{\theta})-K_{n}^{(1/q)}(\underline{%
\theta})\right) \,. 
\end{equation*}
The results of the previous section imply $f_{1}(\theta)=f_{2}(\underline {%
\theta})=0$ for $q=e^{i\pi\nu}$. As an induction assumption we take $f_{n-2}(%
\underline{\theta}^{\prime\prime})=0$. The function $f_{n}(\underline{\theta}%
)$ is meromorphic in terms of the $x_{i}=e^{\theta_{i}}$ with at most simple
poles at $x_{i}=\pm x_{j}$ since $\sinh%
\theta_{ij}=(x_{i}+x_{j})(x_{i}-x_{j})/(2x_{i}x_{j})$. The residues of the
poles at $x_{i}=x_{j}$ vanish because of the symmetry under the exchange of $%
x_{i}\leftrightarrow x_{j}$. The residues at $x_{i}=-x_{j}$ are proportional
to $f_{n-2}(\underline{\theta}^{\prime\prime})$ because of the recursion
relation $(iii)$. Furthermore it can be shown \cite{BK2} that $f_{n}(%
\underline{\theta})\rightarrow0$ for $x_{i}\rightarrow\infty$. Therefore $%
f_{n}(\underline{\theta})$ vanishes identically by Liouville's theorem.

The factor $\frac{\pi\nu}{\sin\pi\nu}$ in (\ref{mass}) modifies the
classical equation and has to be considered as a quantum correction. For the
sin- and sinh-Gordon model an analogous quantum field equation has been
discussed previously \cite{Sm2,MS}. Note that in particular at the `free
fermion point' $\nu\rightarrow1~(\beta^{2}\rightarrow4\pi)$ this factor
diverges, a phenomenon which is to be expected by investigations of the
short distance behavior \cite{ST}. For fixed bare mass square $\alpha$ and $%
\nu \rightarrow2,3,4,\dots$ the physical mass goes to zero. These values of
the coupling are known to be specific: 1) the Bethe Ansatz vacuum in the
language of the massive Thirring model shows phase transitions \cite{Ko} and
2) the model at these points is related \cite{K3,LeC,Sm1} to Baxters
RSOS-models which correspond to minimal conformal models with central charge 
$c=1-6/(\nu(\nu+1))$.

\subsubsection*{The trace of the energy momentum tensor}

As a further operator equation we find \cite{BK1,BK2} (see also \cite{Sm2})
that the trace of the energy momentum tensor satisfies 
\begin{equation}
T_{~\mu}^{\mu}(x)=-2\frac{\alpha}{\beta^{2}}\left( 1-\frac{\beta^{2}}{8\pi }%
\right) \left( :\!\cos\beta\varphi\!:(x)-1\right) .   \label{T}
\end{equation}
Again this operator equation is to be understood as a relation of all its
matrix elements. The equation is modified compared to the classical one by a
quantum correction $(1-\beta^{2}/8\pi)$. As a consequence of this fact the
model will be conformal invariant in the limit $\beta^{2}\rightarrow8\pi$
for fixed bare mass square $\alpha$. This is related to a
Berezinski-Kosterlitz-Thouless \cite{KS} phase transition.

This results may be checked again in perturbation theory by Feynman graph
expansions. The quantum corrections of the trace of the energy momentum
tensor (\ref{T}) yields 
\begin{equation*}
\langle\,p\,|\,\!:\!\cos\beta\varphi\!:(0)-1|\,p\,\rangle=-\beta^{2}\left( 1+%
\frac{\beta^{2}}{8\pi}\right) +O(\beta^{6}). 
\end{equation*}
This again agrees with the exact formula above since the usual normalization
for the energy momentum given by $\int dx\,T^{0\mu}(x)|\,p\,\rangle=p^{\mu
}|\,p\,\rangle$ implies $\langle\,p\,|\,T_{~\mu}^{\mu}|\,p\,\rangle=2m^{2}$.

\section{\ Wightman functions}

As the simplest case we consider the two-point function of two local scalar
operators $\mathcal{O}(x)$ and $\mathcal{O}^{\prime}(x)$%
\begin{equation*}
w(x)=\langle\,0\,|\,\mathcal{O}(x)\,\mathcal{O}^{\prime}(0)\,|\,0\,\rangle. 
\end{equation*}

\subsubsection*{Summation over all intermediate states}

Inserting a complete set of in-states we may write%
\begin{align*}
w(x) & =\sum_{n=0}^{\infty}\frac{1}{n!}\int\frac{dp_{1}}{2\pi2\omega_{1}}%
\dots\int\frac{dp_{n}}{2\pi2\omega_{n}}e^{-ix(p_{1}+\dots+p_{n})} \\
& \times\langle\,0\,|\,\mathcal{O}(0)\,|p_{1},\ldots,p_{n}\,\rangle
^{in}\,^{in}\langle p_{n},\ldots,p_{1}\,|\,\mathcal{O}^{\prime}(0)\,|\,0\,%
\rangle \\
& =\sum_{n=0}^{\infty}\frac{1}{n!}\int d\theta_{1}\dots\int d\theta
_{n}e^{-ix\sum p_{i}}g_{n}(\underline{\theta})\,.
\end{align*}
We have introduced the functions%
\begin{align*}
g_{n}(\underline{\theta}) & =\frac{1}{\left( 4\pi\right) ^{n}}\langle\,0\,|\,%
\mathcal{O}(0)\,|p_{1}, \ldots,p_{n}\,\rangle^{in}\,^{in}\langle
p_{n},\ldots,p_{1}\,|\,\mathcal{O}^{\prime}(0)\,|\,0\,\rangle \\
& =\frac{1}{\left( 4\pi\right) ^{n}}\mathcal{O}(\theta_{1},\ldots
,\theta_{n}\,)\,\mathcal{O}^{\prime}(\theta_{n}+i\pi,\ldots,\theta_{1}+i\pi)%
\,.
\end{align*}
where crossing has been used. In particular we consider exponentials of a
scalar bose field 
\begin{equation*}
\mathcal{O}^{(\prime)}(x)=\,:\!e^{i\gamma^{(\prime)}\varphi(x)}\!: 
\end{equation*}
where $:\dots:$ means normal ordering with respect to the physical vacuum
which means that 
\begin{equation*}
\,\langle\,0\,|:\!e^{i\gamma^{(\prime)}\varphi(x)}\!:|\,0\,\rangle=1 
\end{equation*}
and therefore $g_{0}=1$.

\paragraph{The Log of the two-point function}

For $g_{0}=1$ we may write (see also \cite{Sm2})%
\begin{align*}
w(x) & =1+\sum_{n=1}^{\infty}\frac{1}{n!}\int d\theta_{1}\dots\int
d\theta_{n}e^{-ix\sum p_{i}}g_{n}(\underline{\theta}) \\
& =\exp\sum_{n=1}^{\infty}\frac{1}{n!}\int d\theta_{1}\dots\int d\theta
_{n}e^{-ix\sum p_{i}}h_{n}(\underline{\theta})\,
\end{align*}
It is well known that the functions $g_{n}$ and $h_{n}$ are related by 
\begin{equation*}
g_{I}=\sum_{I_{1}\cup\dots\cup I_{k}=I}h_{I_{1}}\dots h_{I_{k}}\, 
\end{equation*}
where we use the short hand notation $g_{I}=g_{n}(\theta_{1},\dots,\theta
_{n})$ with $I=\left\{ 1,\dots,n\right\} $. The relations of the $g$'s and
the $h$'s may be depicted with ~ $g=%
\begin{picture}(22,12)\put(10,5){%
\oval(20,10)}\end{picture}$ ~and~ $h=%
\begin{picture}(22,12)\put(0,0){%
\framebox(20,10){}}\end{picture}$ ~as 
\begin{equation*}
\unitlength.45mm%
\begin{picture}(240,32)(0,-2) \footnotesize\put(20,20){\oval
(40,20)[]} \put(10,5){\line(0,1){5}} \put(8,-2){$1$} \put(20,5){\makebox
(0,0){$\dots$}} \put(30,5){\line(0,1){5}} \put(28,-2){$n$} \put(47,15){=}%
\put(60,10){\framebox(30,20){}}
\put(65,5){\line(0,1){5}} \put(63,-2){$1$} \put(75,5){\makebox(0,0){$\dots$}%
}%
\put(85,5){\line(0,1){5}} \put(83,-2){$n$} \put(96,15){$+~\displaystyle
\sum_{i=1}^n$} \put(120,10){\framebox(20,15){}} \put(125,0){\line(0,1){10}}%
\put(127.5,4){.\,.} \put(135,0){\line(0,1){10}} \put(150,10){\framebox
(10,10){}}%
\put(155,5){\line(0,1){5}} \put(153,-2){$i$} \put(170,14){$+~\cdots$}%
\put(200,10){\framebox(10,10){}}
\put(205,5){\line(0,1){5}} \put(203,-2){$1$} \put(220,15){\makebox
(0,0){$\dots$}} \put(230,10){\framebox(10,10){}}
\put(235,5){\line(0,1){5}} \put(233,-2){$n$} \end{picture}
\end{equation*}
For example 
\begin{align*}
g_{1} & =h_{1} \\
g_{12} & =h_{12}+h_{1}h_{2} \\
g_{123} & =h_{123}+h_{12}h_{3}+h_{13}h_{2}+h_{23}h_{1}+h_{1}h_{2}h_{3} \\
& \dots
\end{align*}

Due to Lorentz invariance it is sufficient to consider $x=(-i\tau,0)$. Let $%
\mathcal{O}(x)$ and $\,\mathcal{O}^{\prime}(x)$ be scalar operators. Then
the functions $h_{n}(\underline{\theta})$ depend on the rapidity differences
only. We use the formula for the modified Bessel function of the third kind%
\begin{equation*}
i\Delta_{+}(x)=\langle\,0\,|\,\varphi(x)\,\varphi(0)\,|\,0\,\rangle=\frac {1%
}{4\pi}\int d\theta e^{-\tau m\cosh\theta}=\frac{1}{2\pi}K_{0}(m\tau) 
\end{equation*}
to perform one integration 
\begin{align*}
\ln w(x) & =\sum_{n=1}^{\infty}\frac{1}{n!}\int d\theta_{1}\dots\int
d\theta_{n}e^{-\tau m\sum\cosh\theta_{i}}h_{n}(\underline{\theta})\, \\
& =2\sum_{n=1}^{\infty}\frac{1}{n!}\int d\theta_{1}\dots\int d\theta
_{n-1}h_{n}(\theta_{1},\dots,\theta_{n-1},0)K_{0}(m\tau\xi)
\end{align*}

with 
\begin{equation*}
\xi^{2}=\left( \sum_{i=1}^{n-1}\cosh\theta_{i}+1\right) ^{2}-\left(
\sum_{i=1}^{n-1}\sinh\theta_{i}\right) ^{2}. 
\end{equation*}

\subsubsection*{Short distance behavior $\protect\tau\rightarrow0$}

In order to perform the conformal limit of masive models one investigates
the short distance behavior (see e.g.~\cite{Za1,Ca,Sm2,DSC,CF}). For small $%
\tau$ we use the expansion of the modified Bessel function of the third kind
and obtain%
\begin{align*}
\ln w(x) & =-2\sum_{n=1}^{\infty}\frac{1}{n!}\int d\theta_{1}\dots\int
d\theta_{n-1}h_{n}(\theta_{1},\dots,\theta_{n-1},0) \\
& \times\left( \ln m\tau+\ln\xi+\gamma_{E}-\ln2+O\left( \tau^{2}\ln
\tau\right) \right)
\end{align*}
where $\gamma_{E}=0.5772\dots$ is Euler's or Mascheroni's constant.
Therefore the two-point Wightman function has power like behavior for short
distances%
\begin{equation*}
w(x)\approx C\left( m\tau\right) ^{-4\Delta}\quad\text{for }\tau\rightarrow0 
\end{equation*}
where the dimension is given by%
\begin{equation*}
\Delta=\frac{1}{2}\sum_{n=1}^{\infty}\frac{1}{n!}\int d\theta_{1}\dots\int
d\theta_{n-1}h_{n}(\theta_{1},\dots,\theta_{n-1,}0) 
\end{equation*}
if the integrals exist. This is true for the exponentials of bose fields$\,%
\mathcal{O}=:\!e^{i\gamma\varphi(x)}\!:$ because of the asymptotic behavior
for $\func{Re}\theta_{1}\rightarrow\infty$%
\begin{align*}
\mathcal{O}_{n}(\theta_{1,}\theta_{2,}\dots) & =\mathcal{O}_{1}(\theta _{1})%
\mathcal{O}_{n-1}(\theta_{2,}\dots)+O(e^{-\theta_{1}}) \\
g_{n}(\theta_{1},\theta_{2},\dots\theta_{n}) &
=g_{1}g_{n-1}(\theta_{2},\dots\theta_{n})+O(e^{-\left| \theta_{1}\right| })
\end{align*}
as shown above. Therefore the functions $h_{n}$ satisfy (see also \cite{Sm2})%
\begin{equation*}
h_{n}(\underline{\theta})=O(e^{-\left| \theta_{i}\right| })\quad\text{for }%
\func{Re}\theta_{i}\rightarrow\pm\infty\,. 
\end{equation*}
This follows when we distinguish in the relation of the $g$'s and the $h$'s
above the variable $\theta_{1}$ and reorganize the terms on the right hand
side as follows 
\begin{equation*}
g_{I}=\sum_{1\in J\subseteq I}h_{J}g_{I\setminus J}\,. 
\end{equation*}
The constant $C$ is obtained as 
\begin{equation*}
C=\exp\left( -2\sum_{n=1}^{\infty}\frac{1}{n!}\int d\theta_{1}\dots\int
d\theta_{n-1}h_{n}(\theta_{1},\dots,\theta_{n-1},0)\left( \ln\tfrac{1}{2}%
\xi+\gamma_{E}\right) \right) 
\end{equation*}
and it should be related to the vacuum expectation value $G=\,\langle \,0\,|%
\mathcal{O}(x)|\,0\,\rangle_{C}$ in the `conformal normalization' 
\begin{equation*}
C=m^{4\Delta}G^{-2}\,. 
\end{equation*}
Such vacuum expectation value was calculated in \cite{LZ} for the
sine-Gordon model.

\subsubsection*{Examples}

For simplicity let us consider the case that both operator are the
exponential of the field at the special value $\gamma=\gamma^{\prime}=\beta$%
\begin{equation*}
\mathcal{O}(x)=\mathcal{O}^{\prime\dagger}(x)=\,:\!e^{i\beta\varphi(x)}\!: 
\end{equation*}

\paragraph{1. The free case:}

For the free case $h_{1}=\frac{1}{4\pi}\left| \beta\right| ^{2}$ and all $%
h_{n}=0$ for $n>1$%
\begin{align*}
w(x) & \approx e^{-\frac{1}{2\pi}\left( \gamma_{E}-\ln2\right) \left|
\beta\right| ^{2}}(m\tau)^{-\frac{1}{2\pi}\left| \beta\right| ^{2}}\quad%
\text{for }\tau\rightarrow0 \\
\Delta & =\frac{1}{8\pi}\left| \beta\right| ^{2} \\
C & =e^{-\frac{1}{2\pi}\left( \gamma_{E}-\ln2\right) \left| \beta\right|
^{2}}
\end{align*}
which is a well known result.

\paragraph{2. The sinh-Gordon model}

The 2-particle S-matrix is obtained from the sine-Gordon breather S-matrix (%
\ref{2.1}) for imaginary couplings $\beta$ 
\begin{equation*}
S(\theta)=\frac{\sinh\theta+i\sin\pi\nu}{\sinh\theta-i\sin\pi\nu}\quad\text{%
with }-1\leq\nu=\frac{\beta^{2}}{8\pi-\beta^{2}}\leq0\,. 
\end{equation*}
The 2-particle form factor function is given by (\ref{3.6}) for $-1\leq\nu
\leq0$. The sinh Gordon model has the self-dual point%
\begin{equation*}
\nu=-\,\frac{1}{2}~~\mathrm{or}~~\beta^{2}=-4\pi\,. 
\end{equation*}
The dimension of the exponential of the field $\mathcal{O}(x)=\,:\!e^{i\beta
\varphi(x)}\!:$ for the sinh-Gordon model is in the 1- and 1+2-particle
intermediate state approximation (see Fig.~\ref{f1}) 
\begin{align*}
\Delta_{1+2} & =\frac{1}{2}\left( h_{1}+\frac{1}{2!}\int d\theta
\,h_{2}(\theta,0)+\ldots\right) \\
& =-\frac{\sin\pi\nu}{\pi F(i\pi)}+\left( \frac{\sin\pi\nu}{\pi F(i\pi )}%
\right) ^{2}\int_{-\infty}^{\infty}d\theta\,\left( F(\theta)F(-\theta
)-1\right) +\ldots
\end{align*}
The integral may be calculated exactly with the result \cite{BK4}%
\begin{equation*}
-\frac{\pi}{2}\sin\pi\nu F^{2}(i\pi)-\pi\frac{\cos\pi\nu-1}{\sin\pi\nu }%
+2\left( 1-\frac{\pi\nu\cos\pi\nu}{\sin\pi\nu}\right) 
\end{equation*}
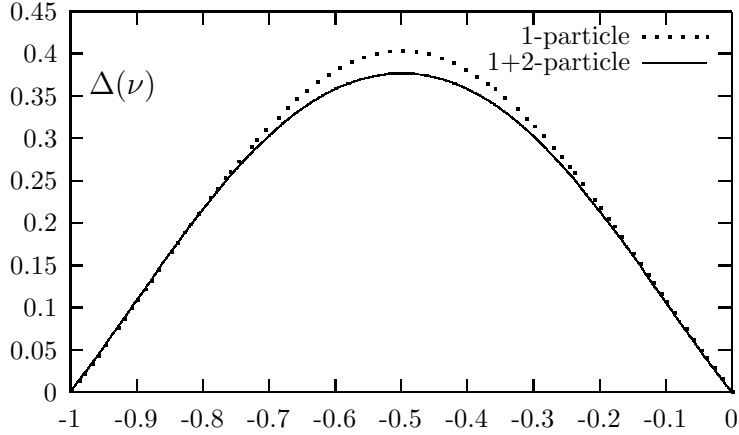
\begin{figure}[h]
\begin{equation*}
\setlength{\unitlength}{0.240900pt} \ifx\plotpoint\undefined%
\newsavebox
{\plotpoint}\fi 
\begin{picture}(1200,720)(0,0)
\font\gnuplot=cmr10 at 10pt
\gnuplot
\sbox{\plotpoint}{\rule[-0.200pt]{0.400pt}{0.400pt}}%
\put(140.0,82.0){\rule[-0.200pt]{4.818pt}{0.400pt}}
\put(120,82){\makebox(0,0)[r]{0}}
\put(1159.0,82.0){\rule[-0.200pt]{4.818pt}{0.400pt}}
\put(140.0,148.0){\rule[-0.200pt]{4.818pt}{0.400pt}}
\put(120,148){\makebox(0,0)[r]{0.05}}
\put(1159.0,148.0){\rule[-0.200pt]{4.818pt}{0.400pt}}
\put(140.0,215.0){\rule[-0.200pt]{4.818pt}{0.400pt}}
\put(120,215){\makebox(0,0)[r]{0.1}}
\put(1159.0,215.0){\rule[-0.200pt]{4.818pt}{0.400pt}}
\put(140.0,281.0){\rule[-0.200pt]{4.818pt}{0.400pt}}
\put(120,281){\makebox(0,0)[r]{0.15}}
\put(1159.0,281.0){\rule[-0.200pt]{4.818pt}{0.400pt}}
\put(140.0,348.0){\rule[-0.200pt]{4.818pt}{0.400pt}}
\put(120,348){\makebox(0,0)[r]{0.2}}
\put(1159.0,348.0){\rule[-0.200pt]{4.818pt}{0.400pt}}
\put(140.0,414.0){\rule[-0.200pt]{4.818pt}{0.400pt}}
\put(120,414){\makebox(0,0)[r]{0.25}}
\put(1159.0,414.0){\rule[-0.200pt]{4.818pt}{0.400pt}}
\put(140.0,481.0){\rule[-0.200pt]{4.818pt}{0.400pt}}
\put(120,481){\makebox(0,0)[r]{0.3}}
\put(1159.0,481.0){\rule[-0.200pt]{4.818pt}{0.400pt}}
\put(140.0,547.0){\rule[-0.200pt]{4.818pt}{0.400pt}}
\put(120,547){\makebox(0,0)[r]{0.35}}
\put(1159.0,547.0){\rule[-0.200pt]{4.818pt}{0.400pt}}
\put(140.0,614.0){\rule[-0.200pt]{4.818pt}{0.400pt}}
\put(120,614){\makebox(0,0)[r]{0.4}}
\put(1159.0,614.0){\rule[-0.200pt]{4.818pt}{0.400pt}}
\put(140.0,680.0){\rule[-0.200pt]{4.818pt}{0.400pt}}
\put(120,680){\makebox(0,0)[r]{0.45}}
\put(1159.0,680.0){\rule[-0.200pt]{4.818pt}{0.400pt}}
\put(140.0,82.0){\rule[-0.200pt]{0.400pt}{4.818pt}}
\put(140,41){\makebox(0,0){-1}}
\put(140.0,660.0){\rule[-0.200pt]{0.400pt}{4.818pt}}
\put(244.0,82.0){\rule[-0.200pt]{0.400pt}{4.818pt}}
\put(244,41){\makebox(0,0){-0.9}}
\put(244.0,660.0){\rule[-0.200pt]{0.400pt}{4.818pt}}
\put(348.0,82.0){\rule[-0.200pt]{0.400pt}{4.818pt}}
\put(348,41){\makebox(0,0){-0.8}}
\put(348.0,660.0){\rule[-0.200pt]{0.400pt}{4.818pt}}
\put(452.0,82.0){\rule[-0.200pt]{0.400pt}{4.818pt}}
\put(452,41){\makebox(0,0){-0.7}}
\put(452.0,660.0){\rule[-0.200pt]{0.400pt}{4.818pt}}
\put(556.0,82.0){\rule[-0.200pt]{0.400pt}{4.818pt}}
\put(556,41){\makebox(0,0){-0.6}}
\put(556.0,660.0){\rule[-0.200pt]{0.400pt}{4.818pt}}
\put(659.0,82.0){\rule[-0.200pt]{0.400pt}{4.818pt}}
\put(659,41){\makebox(0,0){-0.5}}
\put(659.0,660.0){\rule[-0.200pt]{0.400pt}{4.818pt}}
\put(763.0,82.0){\rule[-0.200pt]{0.400pt}{4.818pt}}
\put(763,41){\makebox(0,0){-0.4}}
\put(763.0,660.0){\rule[-0.200pt]{0.400pt}{4.818pt}}
\put(867.0,82.0){\rule[-0.200pt]{0.400pt}{4.818pt}}
\put(867,41){\makebox(0,0){-0.3}}
\put(867.0,660.0){\rule[-0.200pt]{0.400pt}{4.818pt}}
\put(971.0,82.0){\rule[-0.200pt]{0.400pt}{4.818pt}}
\put(971,41){\makebox(0,0){-0.2}}
\put(971.0,660.0){\rule[-0.200pt]{0.400pt}{4.818pt}}
\put(1075.0,82.0){\rule[-0.200pt]{0.400pt}{4.818pt}}
\put(1075,41){\makebox(0,0){-0.1}}
\put(1075.0,660.0){\rule[-0.200pt]{0.400pt}{4.818pt}}
\put(1179.0,82.0){\rule[-0.200pt]{0.400pt}{4.818pt}}
\put(1179,41){\makebox(0,0){0}}
\put(1179.0,660.0){\rule[-0.200pt]{0.400pt}{4.818pt}}
\put(140.0,82.0){\rule[-0.200pt]{250.295pt}{0.400pt}}
\put(1179.0,82.0){\rule[-0.200pt]{0.400pt}{144.058pt}}
\put(140.0,680.0){\rule[-0.200pt]{250.295pt}{0.400pt}}
\put(140.0,82.0){\rule[-0.200pt]{0.400pt}{144.058pt}}
\sbox{\plotpoint}{\rule[-0.500pt]{1.000pt}{1.000pt}}%
\put(1019,640){\makebox(0,0)[r]{1-particle}}
\multiput(1039,640)(20.756,0.000){5}{\usebox{\plotpoint}}
\put(1139,640){\usebox{\plotpoint}}
\put(141,83){\usebox{\plotpoint}}
\multiput(141,83)(12.743,16.383){2}{\usebox{\plotpoint}}
\multiput(162,110)(12.453,16.604){2}{\usebox{\plotpoint}}
\multiput(183,138)(11.784,17.086){2}{\usebox{\plotpoint}}
\put(214.53,183.47){\usebox{\plotpoint}}
\multiput(224,197)(12.173,16.811){2}{\usebox{\plotpoint}}
\multiput(245,226)(11.902,17.004){2}{\usebox{\plotpoint}}
\multiput(266,256)(11.513,17.270){2}{\usebox{\plotpoint}}
\put(297.63,302.62){\usebox{\plotpoint}}
\multiput(307,316)(12.173,16.811){2}{\usebox{\plotpoint}}
\multiput(328,345)(12.173,16.811){2}{\usebox{\plotpoint}}
\put(358.72,386.50){\usebox{\plotpoint}}
\multiput(370,401)(12.354,16.678){2}{\usebox{\plotpoint}}
\multiput(390,428)(13.041,16.147){2}{\usebox{\plotpoint}}
\put(423.11,467.84){\usebox{\plotpoint}}
\multiput(432,478)(13.995,15.328){2}{\usebox{\plotpoint}}
\put(465.46,513.46){\usebox{\plotpoint}}
\put(480.13,528.13){\usebox{\plotpoint}}
\multiput(494,542)(16.132,13.059){2}{\usebox{\plotpoint}}
\put(527.44,568.48){\usebox{\plotpoint}}
\put(544.50,580.26){\usebox{\plotpoint}}
\put(562.30,590.92){\usebox{\plotpoint}}
\put(580.72,600.42){\usebox{\plotpoint}}
\put(600.18,607.62){\usebox{\plotpoint}}
\put(620.16,613.22){\usebox{\plotpoint}}
\put(640.56,617.03){\usebox{\plotpoint}}
\put(661.29,617.99){\usebox{\plotpoint}}
\put(682.00,616.81){\usebox{\plotpoint}}
\multiput(702,613)(19.690,-6.563){2}{\usebox{\plotpoint}}
\put(741.48,598.96){\usebox{\plotpoint}}
\put(759.48,588.71){\usebox{\plotpoint}}
\put(777.17,577.85){\usebox{\plotpoint}}
\put(794.39,566.29){\usebox{\plotpoint}}
\multiput(806,558)(15.759,-13.508){2}{\usebox{\plotpoint}}
\put(841.73,525.97){\usebox{\plotpoint}}
\put(856.35,511.24){\usebox{\plotpoint}}
\multiput(868,499)(13.995,-15.328){2}{\usebox{\plotpoint}}
\put(898.15,465.11){\usebox{\plotpoint}}
\multiput(910,451)(13.350,-15.893){2}{\usebox{\plotpoint}}
\multiput(931,426)(12.354,-16.678){2}{\usebox{\plotpoint}}
\put(962.46,383.72){\usebox{\plotpoint}}
\multiput(972,371)(12.173,-16.811){2}{\usebox{\plotpoint}}
\multiput(993,342)(12.173,-16.811){2}{\usebox{\plotpoint}}
\put(1023.33,299.67){\usebox{\plotpoint}}
\multiput(1035,283)(11.513,-17.270){2}{\usebox{\plotpoint}}
\multiput(1055,253)(11.902,-17.004){2}{\usebox{\plotpoint}}
\multiput(1076,223)(12.173,-16.811){2}{\usebox{\plotpoint}}
\put(1106.44,180.52){\usebox{\plotpoint}}
\multiput(1118,164)(12.064,-16.889){2}{\usebox{\plotpoint}}
\multiput(1138,136)(12.551,-16.531){3}{\usebox{\plotpoint}}
\put(1179,82){\usebox{\plotpoint}}
\sbox{\plotpoint}{\rule[-0.200pt]{0.400pt}{0.400pt}}%
\put(1019,599){\makebox(0,0)[r]{1+2-particle}}
\put(1039.0,599.0){\rule[-0.200pt]{24.090pt}{0.400pt}}
\put(141,83){\usebox{\plotpoint}}
\multiput(141.58,83.00)(0.496,0.643){39}{\rule{0.119pt}{0.614pt}}
\multiput(140.17,83.00)(21.000,25.725){2}{\rule{0.400pt}{0.307pt}}
\multiput(162.58,110.00)(0.496,0.668){39}{\rule{0.119pt}{0.633pt}}
\multiput(161.17,110.00)(21.000,26.685){2}{\rule{0.400pt}{0.317pt}}
\multiput(183.58,138.00)(0.496,0.727){37}{\rule{0.119pt}{0.680pt}}
\multiput(182.17,138.00)(20.000,27.589){2}{\rule{0.400pt}{0.340pt}}
\multiput(203.58,167.00)(0.496,0.692){39}{\rule{0.119pt}{0.652pt}}
\multiput(202.17,167.00)(21.000,27.646){2}{\rule{0.400pt}{0.326pt}}
\multiput(224.58,196.00)(0.496,0.716){39}{\rule{0.119pt}{0.671pt}}
\multiput(223.17,196.00)(21.000,28.606){2}{\rule{0.400pt}{0.336pt}}
\multiput(245.58,226.00)(0.496,0.692){39}{\rule{0.119pt}{0.652pt}}
\multiput(244.17,226.00)(21.000,27.646){2}{\rule{0.400pt}{0.326pt}}
\multiput(266.58,255.00)(0.496,0.753){37}{\rule{0.119pt}{0.700pt}}
\multiput(265.17,255.00)(20.000,28.547){2}{\rule{0.400pt}{0.350pt}}
\multiput(286.58,285.00)(0.496,0.668){39}{\rule{0.119pt}{0.633pt}}
\multiput(285.17,285.00)(21.000,26.685){2}{\rule{0.400pt}{0.317pt}}
\multiput(307.58,313.00)(0.496,0.692){39}{\rule{0.119pt}{0.652pt}}
\multiput(306.17,313.00)(21.000,27.646){2}{\rule{0.400pt}{0.326pt}}
\multiput(328.58,342.00)(0.496,0.643){39}{\rule{0.119pt}{0.614pt}}
\multiput(327.17,342.00)(21.000,25.725){2}{\rule{0.400pt}{0.307pt}}
\multiput(349.58,369.00)(0.496,0.619){39}{\rule{0.119pt}{0.595pt}}
\multiput(348.17,369.00)(21.000,24.765){2}{\rule{0.400pt}{0.298pt}}
\multiput(370.58,395.00)(0.496,0.625){37}{\rule{0.119pt}{0.600pt}}
\multiput(369.17,395.00)(20.000,23.755){2}{\rule{0.400pt}{0.300pt}}
\multiput(390.58,420.00)(0.496,0.546){39}{\rule{0.119pt}{0.538pt}}
\multiput(389.17,420.00)(21.000,21.883){2}{\rule{0.400pt}{0.269pt}}
\multiput(411.58,443.00)(0.496,0.522){39}{\rule{0.119pt}{0.519pt}}
\multiput(410.17,443.00)(21.000,20.923){2}{\rule{0.400pt}{0.260pt}}
\multiput(432.00,465.58)(0.523,0.496){37}{\rule{0.520pt}{0.119pt}}
\multiput(432.00,464.17)(19.921,20.000){2}{\rule{0.260pt}{0.400pt}}
\multiput(453.00,485.58)(0.551,0.495){35}{\rule{0.542pt}{0.119pt}}
\multiput(453.00,484.17)(19.875,19.000){2}{\rule{0.271pt}{0.400pt}}
\multiput(474.00,504.58)(0.625,0.494){29}{\rule{0.600pt}{0.119pt}}
\multiput(474.00,503.17)(18.755,16.000){2}{\rule{0.300pt}{0.400pt}}
\multiput(494.00,520.58)(0.702,0.494){27}{\rule{0.660pt}{0.119pt}}
\multiput(494.00,519.17)(19.630,15.000){2}{\rule{0.330pt}{0.400pt}}
\multiput(515.00,535.58)(0.814,0.493){23}{\rule{0.746pt}{0.119pt}}
\multiput(515.00,534.17)(19.451,13.000){2}{\rule{0.373pt}{0.400pt}}
\multiput(536.00,548.58)(0.967,0.492){19}{\rule{0.864pt}{0.118pt}}
\multiput(536.00,547.17)(19.207,11.000){2}{\rule{0.432pt}{0.400pt}}
\multiput(557.00,559.59)(1.286,0.488){13}{\rule{1.100pt}{0.117pt}}
\multiput(557.00,558.17)(17.717,8.000){2}{\rule{0.550pt}{0.400pt}}
\multiput(577.00,567.59)(1.560,0.485){11}{\rule{1.300pt}{0.117pt}}
\multiput(577.00,566.17)(18.302,7.000){2}{\rule{0.650pt}{0.400pt}}
\multiput(598.00,574.59)(2.269,0.477){7}{\rule{1.780pt}{0.115pt}}
\multiput(598.00,573.17)(17.306,5.000){2}{\rule{0.890pt}{0.400pt}}
\multiput(619.00,579.61)(4.481,0.447){3}{\rule{2.900pt}{0.108pt}}
\multiput(619.00,578.17)(14.981,3.000){2}{\rule{1.450pt}{0.400pt}}
\put(640,581.67){\rule{5.059pt}{0.400pt}}
\multiput(640.00,581.17)(10.500,1.000){2}{\rule{2.529pt}{0.400pt}}
\put(661,581.67){\rule{4.818pt}{0.400pt}}
\multiput(661.00,582.17)(10.000,-1.000){2}{\rule{2.409pt}{0.400pt}}
\multiput(681.00,580.95)(4.481,-0.447){3}{\rule{2.900pt}{0.108pt}}
\multiput(681.00,581.17)(14.981,-3.000){2}{\rule{1.450pt}{0.400pt}}
\multiput(702.00,577.93)(2.269,-0.477){7}{\rule{1.780pt}{0.115pt}}
\multiput(702.00,578.17)(17.306,-5.000){2}{\rule{0.890pt}{0.400pt}}
\multiput(723.00,572.93)(1.560,-0.485){11}{\rule{1.300pt}{0.117pt}}
\multiput(723.00,573.17)(18.302,-7.000){2}{\rule{0.650pt}{0.400pt}}
\multiput(744.00,565.93)(1.135,-0.489){15}{\rule{0.989pt}{0.118pt}}
\multiput(744.00,566.17)(17.948,-9.000){2}{\rule{0.494pt}{0.400pt}}
\multiput(764.00,556.92)(0.967,-0.492){19}{\rule{0.864pt}{0.118pt}}
\multiput(764.00,557.17)(19.207,-11.000){2}{\rule{0.432pt}{0.400pt}}
\multiput(785.00,545.92)(0.814,-0.493){23}{\rule{0.746pt}{0.119pt}}
\multiput(785.00,546.17)(19.451,-13.000){2}{\rule{0.373pt}{0.400pt}}
\multiput(806.00,532.92)(0.702,-0.494){27}{\rule{0.660pt}{0.119pt}}
\multiput(806.00,533.17)(19.630,-15.000){2}{\rule{0.330pt}{0.400pt}}
\multiput(827.00,517.92)(0.618,-0.495){31}{\rule{0.594pt}{0.119pt}}
\multiput(827.00,518.17)(19.767,-17.000){2}{\rule{0.297pt}{0.400pt}}
\multiput(848.00,500.92)(0.554,-0.495){33}{\rule{0.544pt}{0.119pt}}
\multiput(848.00,501.17)(18.870,-18.000){2}{\rule{0.272pt}{0.400pt}}
\multiput(868.00,482.92)(0.498,-0.496){39}{\rule{0.500pt}{0.119pt}}
\multiput(868.00,483.17)(19.962,-21.000){2}{\rule{0.250pt}{0.400pt}}
\multiput(889.58,460.85)(0.496,-0.522){39}{\rule{0.119pt}{0.519pt}}
\multiput(888.17,461.92)(21.000,-20.923){2}{\rule{0.400pt}{0.260pt}}
\multiput(910.58,438.77)(0.496,-0.546){39}{\rule{0.119pt}{0.538pt}}
\multiput(909.17,439.88)(21.000,-21.883){2}{\rule{0.400pt}{0.269pt}}
\multiput(931.58,415.43)(0.496,-0.651){37}{\rule{0.119pt}{0.620pt}}
\multiput(930.17,416.71)(20.000,-24.713){2}{\rule{0.400pt}{0.310pt}}
\multiput(951.58,389.53)(0.496,-0.619){39}{\rule{0.119pt}{0.595pt}}
\multiput(950.17,390.76)(21.000,-24.765){2}{\rule{0.400pt}{0.298pt}}
\multiput(972.58,363.45)(0.496,-0.643){39}{\rule{0.119pt}{0.614pt}}
\multiput(971.17,364.73)(21.000,-25.725){2}{\rule{0.400pt}{0.307pt}}
\multiput(993.58,336.37)(0.496,-0.668){39}{\rule{0.119pt}{0.633pt}}
\multiput(992.17,337.69)(21.000,-26.685){2}{\rule{0.400pt}{0.317pt}}
\multiput(1014.58,308.29)(0.496,-0.692){39}{\rule{0.119pt}{0.652pt}}
\multiput(1013.17,309.65)(21.000,-27.646){2}{\rule{0.400pt}{0.326pt}}
\multiput(1035.58,279.09)(0.496,-0.753){37}{\rule{0.119pt}{0.700pt}}
\multiput(1034.17,280.55)(20.000,-28.547){2}{\rule{0.400pt}{0.350pt}}
\multiput(1055.58,249.29)(0.496,-0.692){39}{\rule{0.119pt}{0.652pt}}
\multiput(1054.17,250.65)(21.000,-27.646){2}{\rule{0.400pt}{0.326pt}}
\multiput(1076.58,220.21)(0.496,-0.716){39}{\rule{0.119pt}{0.671pt}}
\multiput(1075.17,221.61)(21.000,-28.606){2}{\rule{0.400pt}{0.336pt}}
\multiput(1097.58,190.29)(0.496,-0.692){39}{\rule{0.119pt}{0.652pt}}
\multiput(1096.17,191.65)(21.000,-27.646){2}{\rule{0.400pt}{0.326pt}}
\multiput(1118.58,161.26)(0.496,-0.702){37}{\rule{0.119pt}{0.660pt}}
\multiput(1117.17,162.63)(20.000,-26.630){2}{\rule{0.400pt}{0.330pt}}
\multiput(1138.58,133.40)(0.498,-0.659){79}{\rule{0.120pt}{0.627pt}}
\multiput(1137.17,134.70)(41.000,-52.699){2}{\rule{0.400pt}{0.313pt}}
\put(170,550){$\Delta(\nu)$}
\end{picture}
\end{equation*}
\caption{\emph{Dimension of an exponential of the field for the sinh-Gordon
model: 1- and 1+2-particle intermediate state contributions}}
\label{f1}
\end{figure}
In principle the higher particle intermediate state integrals may also be
calculated, however, up to now we could not derive a general formula. This
is possible for the scaling Ising model which is considered as the next
example. The constant $C$ in the approximation of 1-intermediate states is
given as%
\begin{equation*}
C_{1}=\exp\left( -2h_{1}\left( \gamma_{E}-\ln2\right) \right) =\exp\left( 4%
\frac{\sin\pi\nu}{\pi F(i\pi)}\left( \gamma_{E}-\ln2\right) \right) 
\end{equation*}
We have not calculated the integrals appearing in higher-particle
intermediate state contributions.

\paragraph{3. The scaling Ising model}

\cite{BKW,CM,YZ,CF} The S-matrix is $S=-1$. The form factors of the order
parameter $\sigma$ are non-vanishing for odd $n$ \cite{BKW} 
\begin{equation*}
\langle\,0\,|\,\sigma(0)\,|p_{1},\ldots,p_{n}\,\rangle^{in}=\left( 2i\right)
^{\frac{n-1}{2}}\prod_{i<j}\tanh\frac{1}{2}\theta_{ij}\, 
\end{equation*}
and for the disorder parameter $\mu$ for even $n$ 
\begin{equation*}
\langle\,0\,|\,\mu(0)\,|p_{1},\ldots,p_{n}\,\rangle^{in}=\left( 2i\right) ^{%
\frac{n}{2}}\prod_{i<j}\tanh\frac{1}{2}\theta_{ij}\,. 
\end{equation*}
We introduce the operator $\mathcal{O}(x)=\mu(x)+\sqrt{2i}\sigma(x)$
normalized such that $\mathcal{O}_{0}=1$ and $\mathcal{O}_{1}=\sqrt{2i}.$ It
obviously satisfies the asymptotic cluster behavior like an exponential of a
bose field 
\begin{equation*}
\mathcal{O}_{n}(\theta_{1},\dots,\theta_{n})=\mathcal{O}_{1}(\theta _{1})%
\mathcal{O}_{n-1}(\theta_{2},\dots,\theta_{n})+O(e^{-\theta_{1}})\quad\text{%
for }\theta_{1}\rightarrow\infty\,. 
\end{equation*}
This allows us to apply the same methods as above 
\begin{align*}
\langle\,0\,|\,\mathcal{O}(x)\,\mathcal{O}^{\dagger}(x)\,|\,0\,\rangle &
=\sum_{n=1}^{\infty}\frac{1}{n!}\int d\theta_{1}\dots\int d\theta
_{n}e^{-ix\sum p_{i}}g_{n}(\underline{\theta}) \\
& \approx C\left( m\tau\right) ^{-4\Delta}\quad\text{for }\tau\rightarrow0
\end{align*}
with $g_{n}(\underline{\theta})=(2\pi)^{-n}\prod_{i<j}\tanh^{2}\frac{1}{2}%
\theta_{ij}$. To obtain dimension the $\Delta$ and the constant $C$ we have
to calculate 
\begin{align*}
\Delta & =\frac{1}{2}\sum_{n=1}^{\infty}\frac{1}{n!}\int
d\theta_{1}\dots\int d\theta_{n-1}h_{n}(\theta_{1},\dots,\theta_{n-1,}0)=%
\frac{1}{2}\sum _{n=1}^{\infty}\frac{1}{n!}\left( \frac{1}{2\pi}\right)
^{n}I_{n} \\
C & =\exp\left( -2\sum_{n=1}^{\infty}\frac{1}{n!}\int d\theta_{1}\dots\int
d\theta_{n-1}h_{n}(\theta_{1},\dots,\theta_{n-1},0)\left( \ln\xi+\gamma
_{E}-\ln2\right) \right)
\end{align*}
All integrals in the sum over the intermediate states in the formula of the
dimension can be performed. The result may be expressed by a recursion
relation%
\begin{equation*}
\begin{array}{lll}
I_{1} & = & 1 \\ 
I_{2} & = & \int d\theta\left( \tanh^{2}\theta-1\right) =-4 \\ 
I_{3} & = & \int d\theta_{1}\int d\theta_{2}\left\{ \tanh^{2}\tfrac{1}{2}%
\theta_{12}\tanh^{2}\tfrac{1}{2}\theta_{1}\tanh^{2}\tfrac{1}{2}\theta
_{2}\right. \\ 
&  & \left. -\tanh^{2}\tfrac{1}{2}\theta_{12}-\tanh^{2}\tfrac{1}{2}\theta
_{1}-\tanh^{2}\tfrac{1}{2}\theta_{2}+2\right\} =4\pi^{2} \\ 
I_{n} & = & \left( n-2\right) ^{2}I_{3}I_{n-2}%
\end{array}
\end{equation*}
The sum over all intermediate states can be performed by solving the
recursion relation which yields 
\begin{align*}
\Delta & =\frac{1}{4\pi}\sum_{k=0}^{\infty}\frac{\Gamma^{2}\left(
2k+1\right) }{\Gamma^{2}\left( k+1\right) \Gamma\left( 2k+2\right) }4^{-k}-%
\frac{1}{8\pi^{2}}\sum_{k=1}^{\infty}\frac{\Gamma^{2}\left( k\right) }{%
\Gamma\left( 2k+1\right) }4^{k} \\
& =\frac{1}{8}-\frac{1}{16}=\frac{1}{16}
\end{align*}
and which is a well known result (see for example \cite{B-J}). For the
constant $C$ we calculated the 1- and 2-particle intermediate state
contributions%
\begin{align*}
C_{1+2} & =\exp\left( -\frac{1}{\pi}\left( \gamma_{E}-\ln2\right) \right) \\
& \times\exp\left( -\frac{1}{\left( 2\pi\right) ^{2}}\int d\theta\left(
\tanh^{2}\tfrac{1}{2}\theta-1\right) \left( \ln\cosh\tfrac{1}{2}%
\theta+\gamma_{E}\right) \right) \\
& =e^{-\left( \gamma_{E}-\ln2\right) /\pi}e^{-\left( \ln2-\gamma
_{E}-1\right) /\pi^{2}}=1.\,134\,8\,.
\end{align*}
Using 
\begin{align*}
\langle\,0\,|\,\mathcal{O}(x)\,\mathcal{O}^{\dagger}(x)\,|\,0\,\rangle &
=\langle\,0\,|\,\left( \mu+\sqrt{2i}\sigma\right) (x)\,\left( \mu+\sqrt {2i}%
\sigma\right) ^{\dagger}(0)\,|\,0\,\rangle \\
& \approx\left( C^{(\mu)}+2C^{(\sigma)}\right) \left( m\tau\right)
^{-4\Delta}
\end{align*}
this approximation may be compared with the results of \cite{B-J} 
\begin{equation*}
C=C^{(\mu)}+2C^{(\sigma)}=1.\,084\,8\,. 
\end{equation*}

\section*{Acknowledgments}

We thank A. Belavin, M. Jimbo, V.A. Fateev, A. Fring, T. Miwa, Y. Pugai,
F.A. Smirnov, R. Schrader, B. Schroer, J. Teschner, and A.B. Zamolodchikov
for discussions. H.B. was supported by DFG, Sonderforschungsbereich 288
`Differentialgeometrie und Quantenphysik' and partially by the grants INTAS
99-01459 and INTAS 00-561.
This work is also supported by the EU network EUCLID, 'Integrable models and
applications: from strings to condensed matter', HPRN-CT-2002-00325.

\end{document}